\documentclass[journal,hideappendix]{vgtc}        
\usepackage{amsmath}
\usepackage{array}
\usepackage{xcolor}
\usepackage{tabularx}
\usepackage{enumitem}
\definecolor{NavyBlue}{RGB}{0,0,128}

\PassOptionsToPackage{dvipsnames,svgnames}{xcolor}

\onlineid{0}

\vgtccategory{Research}

\vgtcpapertype{theory/model}

\title{Building and Eroding: Exogenous and Endogenous Factors\\ that Influence Subjective Trust in Visualization}

\author{
  \authororcid{R. Jordan Crouser}{0000-0001-9936-0791},
  Syrine Matoussi,
  Lan Kung, 
  Saugat Pandey,
  Oen G. McKinley, and 
  \authororcid{Alvitta Ottley}{0000-0002-9485-276X}
}

\authorfooter{
  \item
  	R. Jordan Crouser, Syrine Matoussi, and Lan Kung are with the\\ Human Computation \& Visualization Laboratory at Smith College.\\
  	E-mail: \{jcrouser,cmatoussi,ekung\}@smith.edu
  \item
  	Saugat Pandey,  Oen G. McKinley, and Alvitta Ottley are with the Visual Interface and Behavior Exploration (VIBE) Laboratory at Washington University in St. Louis.
  	E-mail: \{p.saugat,m.oen,alvitta\}@wustl.edu.
}
\abstract{%
  Trust is a subjective yet fundamental component of human-computer interaction, and is a determining factor in shaping the efficacy of data visualizations. Prior research has identified five dimensions of trust assessment in visualizations (credibility, clarity, reliability, familiarity, and confidence), and observed that these dimensions tend to vary predictably along with certain features of the visualization being evaluated. This raises a further question: how do the design features driving viewers' trust assessment vary with the characteristics of the viewers themselves? By reanalyzing data from these studies through the lens of individual differences, we build a more detailed map of the relationships between design features, individual characteristics, and trust behaviors. In particular, we model the distinct contributions of \textbf{endogenous} design features (such as visualization type, or the use of color) and \textbf{exogenous} user characteristics (such as visualization literacy), as well as the interactions between them. We then use these findings to make recommendations for individualized and adaptive visualization design.
}

\keywords{Trust, data visualization, individual differences, personality}

\graphicspath{{figs/}{figures/}{pictures/}{images/}{./}} 

\usepackage{tabu}                      
\usepackage{booktabs}                  
\usepackage{lipsum}                    
\usepackage{mwe}                       

\usepackage{mathptmx}                  

\begin{document}


\firstsection{Introduction}

\maketitle


In today's information age, trust plays a critical role in influencing decision-making processes across different fields. Furthermore, the widespread dissemination of misinformation and disinformation has highlighted the urgent need to assist individuals in distinguishing truth from falsehood, presenting a significant societal challenge.
As data visualizations continue to become indispensable tools for conveying complex information in accessible forms, 
it is important to acknowledge that they, too, are susceptible to manipulation, distortion, and misinterpretation like any other mode of communication. Thus, understanding the factors that influence people's trust in specific data visualizations is critical for designing effective reasoning aides.

As individuals engage with data visualizations, they are implicitly working to evaluate the accuracy and credibility of the presented information in order to inform their judgments and actions. Trust inspires confidence, and when the data underlying the visualization is sound, this confidence can support well-informed decision-making. However, trust doesn't exist by itself -- it is relational. The same degree of trust in a \emph{misleading} visualization or mistrust in a faithful or unbiased visualization can propagate erroneous conclusions and misguided actions. 

Furthermore, the significance of trust in visualizations and technology extends to its profound implications for public perception and engagement across various sectors. In contexts such as organizational integration of technologies like artificial intelligence~\cite{hoffman2013trust} and web services security ~\cite{ratnasingam2002importance}, trust is critical for user acceptance and adoption~\cite{bahmanziari2003trust}. In domains such as journalism, science communication, and public health, visualizations serve as powerful tools for engaging audiences and conveying complex information. However, the effectiveness of these visualizations hinges on their perceived trustworthiness ~\cite{elhamdadi2023vistrust}. A nuanced understanding of the factors that shape trust can empower practitioners to craft visualizations that resonate with their audiences, thereby fostering heightened engagement and comprehension.

Pandey et al. conducted two studies to explore the relationships between various visual design features and five interrelated facets of trust: credibility, clarity, reliability, familiarity, and confidence~\cite{pandey2023you}. The first study asked participants to rate various visualizations along these dimensions and identified several design features that have a significant correlation with participants' subjective perceptions of trust. They observed that colorful visualizations and visual embellishments garnered greater favor among participants. Moreover, visualizations from news media were perceived as \emph{more credible} and \emph{more reliable} than those from scientific or governmental agencies, even in cases where information regarding the source of the visualization was not explicitly available. This suggests that participants may be picking up on disciplinary norms around data-driven communication. It has been hypothesized that scientific and government entities' tendency toward technical and data-dense designs may render them less accessible (and therefore less trustworthy) to everyday viewers. 

The second experiment investigated how individuals weigh these trust dimensions within the context of visualization design. First, stimuli from the first experiment were sampled to retain visualizations that had both high response rates and low variance and then were further down-sampled to only those examples with the highest and lowest scores along each dimension. Participants were then asked to assess each example along the original 5 dimensions, as well as their \emph{overall trust} in the visualization. In this experiment, factors such as source credibility, content familiarity, and type of visualization emerged as significant correlates with overall trust rankings. 

These findings underscore the complex interplay between visualization design and perceptions of trust. Moreover, we know that patterns of interactions with data visualizations are not universal, but are instead modulated by individual differences between users~\cite{liu2020survey}. Building on these insights, this paper explores the relationships between factors \emph{endogenous} to the visualization (i.e. visual metaphor, use of color, source, etc.) and \emph{exogenous} factors (such as individual differences in personality or cognitive ability, educational background, and cultural influences), and how these factors combine to affect perceived trust. Striking a balance among these factors is paramount for effectively communicating information, fostering accurate comprehension, and helping decision-making across diverse audiences. 

\subsection{Contributions}
This work makes the following contributions to the study of trust in visualization:
\begin{enumerate}
\item We conducted a supplemental reanalysis of data collected by Pandey et al.~\cite{pandey2023you} through the lens of \textbf{individual differences}.
\item We identified \emph{visualization type} and \emph{visualization literacy} as key \textbf{endogenous} and \textbf{exogenous} predictors of trust, respectively.
\item We observed that endogenous and exogenous factors have\\ \textbf{nontrivial interactions} in how they influence trust. 
\end{enumerate}
By considering these factors, we aim to deepen our understanding of the nuanced dynamics between visualization design and trust perception.

\section{Background}

Prior work has identified a variety of factors that can affect how users perceive data visualizations. Some of these factors are endogenous to the visualization itself such as human recognizable objects, visualization type, data-ink ratio, visual density, and color. Others are exogenous contextualizing the user's experience of the visualization through subjective and socially mediated features. However, a standard approach to evaluating trustworthiness is yet to be defined~\cite{borgo2020development}. 

\subsection{Endogenous Factors}
Relevant textual and visual elements can help viewers retain important messages and trends, making visualizations more memorable~\cite{borkin2015beyond}. Memorability is closely intertwined with comprehension and clarity, dimensions by which we measure trust. Thus, when crafting clear and comprehensible visualizations, it is crucial to consider the specific visual features and affordances offered by the design.

\subsubsection{Visualization Type} 

\begin{figure}[b!]
\centering
    \includegraphics[width=0.45\textwidth]{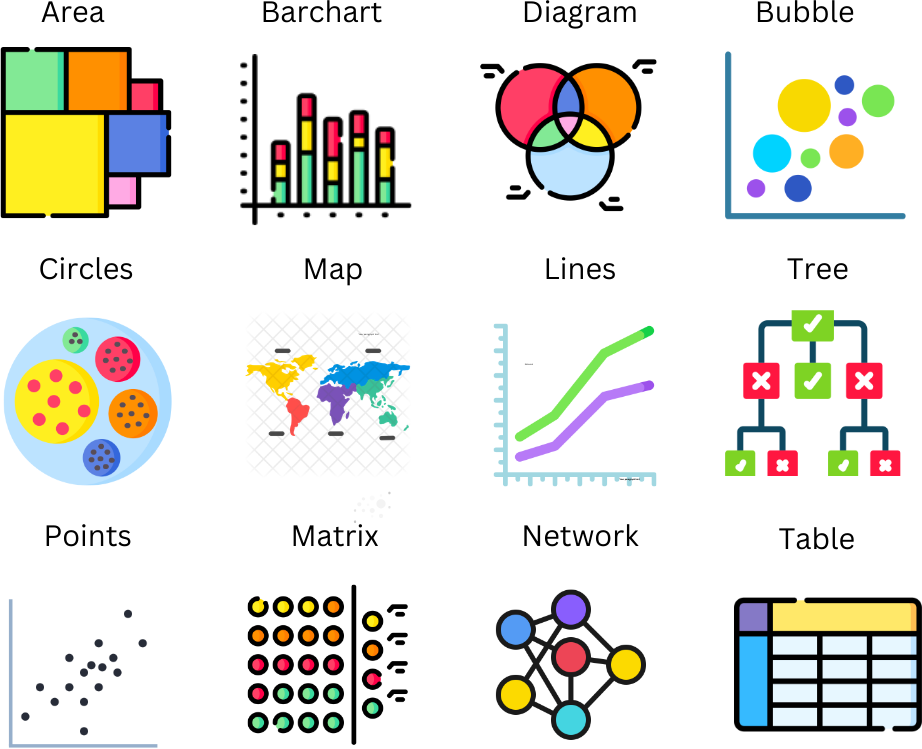}
    \caption{Visualization Types}
    \label{fig:vistypes}
\end{figure}

In their early experiments in the 1980s, Cleveland and McGill~\cite{cleveland1984graphical} observed that humans excel at interpreting length in visualizations (e.g., bar charts), but are less precise with direction (e.g., line charts and slope graphs), and even less accurate with angles (e.g., pie charts). Area (e.g., bubble graphs), volume (e.g., 3D graphs), and curvature (e.g., donut charts) can be particularly challenging. These observations have been used to generate visualization hierarchies, taxonomies, and even phylogenies~\cite{li2015exploring} which break the space of visualizations into broad categories. 
Of course, the clarity of each visual mapping is contingent upon data content and organizational context. For example, line graphs effectively illustrate changes over time, while scatterplots are adept at revealing trends, outliers, and density\cite{magnuson2016data}. Diverging bar charts were favored over clustered ones to improve comprehension in healthcare data \cite{evergreen2019effective}. The previous study by Pandey et al.~\cite{pandey2023you} identified clarity as a core component of trust formation, and so in this supplemental analysis we will also consider the influence of visualization type. To do so, we will adopt the taxonomy presented by Borkin et. al~\cite{borkin2013makes} for classifying static data visualizations (see Fig.~\ref{fig:vistypes}). 

\subsubsection{Visual Embellishments}
Several studies show the impact of visual features on memorability and understanding. An analysis of the MassVis dataset~\cite{borkin2015beyond} found that human-recognizable objects within visualizations can attract attention and aid recognition while compelling titles and appropriate data-ink ratios enhance message conveyance. Interestingly, visualizations with low data-to-ink ratios and high visual densities, often associated with clutter, were found to be more memorable than minimal, clean visualizations~\cite{borkin2013makes}. Unexpected and relevant visual enhancements can boost the recall of thematic elements ~\cite{pena2020memorability}. In fact, unique visualization types like pictorial representations, grids/matrices, trees and networks, and diagrams tend to be more memorable than common graphs like circles, areas, points, bars, and lines. Conversely, unrelated embellishments may distract or hinder memorability when viewers fail to perceive a meaningful connection between informational content and design modifications. 

\subsubsection{Color} Investigations into the impact of size variation in visual elements offer practical guidance for designers in selecting colors that optimize visual contrast~\cite{szafir2017modeling}. Elongated shapes, such as bars and lines, amplify the distinctiveness of colors compared to fixed-thickness shapes. This heightened distinctiveness plays a crucial role in enhancing memory and comprehension, thus reinforcing the significance of color selection in design. Moreover, color not only influences visual contrast but also affects statistical judgments, thereby shaping the understanding process of graph information. Color perception is therefore pivotal in visualization design, as comprehension is a key dimension of trust.

\subsection{Exogenous Features}

In addition to studying the visualization itself, a growing body of literature has begun to explore the role of personality traits, cognitive characteristics, and behavioral variance in modulating individuals' comprehension and interpretation of visualizations~\cite{liu2020survey}. In this section, we delve deeper into the influence of the viewer's features on their perceptions of visualizations.

\subsubsection{Individual Differences}

Psychological research has demonstrated the profound impact of personality traits and cognitive abilities on problem-solving approaches and behavioral patterns~\cite{liu2020survey}. These disparities manifest distinctly across various tasks and contexts. In the realm of computational sciences, there is a growing acknowledgment of how these individual differences influence human-machine interactions ~\cite{aykin1991individual}. Within the data visualization community factors like extroversion, neuroticism, openness to experience, spatial ability, perceptual speed, and memory have been identified as factors that can affect visualization use ~\cite{liu2020survey}. For example, verbal working memory is believed to impact the processing of textual components within visualizations, such as labels, legends, task descriptions, and accompanying texts.  And personality features such as locus of control significantly correlate with speed, accuracy, and strategic approach.
    
\subsubsection{Visualization Literacy} 

The term refers to the capacity and proficiency in deciphering and comprehending visually depicted data to derive insights. It significantly impacts the comprehension of graphs. Visualizer-verbalizer cognitive style, numeracy, and need for cognition are key factors influencing visualization literacy due to the prevalence of numerical data in graphs and the significance of perceptual and cognitive operations in extracting insights from data visualizations ~\cite{lee2019correlation}.
To measure data literacy in this work, we use the Mini-VLAT, a brief but practical visualization literacy test ~\cite{pandey2023mini}, consisting of a 12-item short-form abbreviated version of the 53-item Visualization Literacy Assessment Test (VLAT).

\subsubsection{Education}
Education level has been linked to both generalized social trust~\cite{hooghe2012cognitive} and domain-specific trust in areas such as politics~\cite{ugur2020does}. We hypothesize that education, particularly in fields related to statistics, data science, or critical thinking, may help to equip individuals with the skills required to critically evaluate the reliability and relevance of visualized data. More highly educated individuals may be more likely to question the sources of data, the methods used for data collection and analysis, and the appropriateness of the visualization techniques used. On the other hand, those with less exposure to formal education in these areas might either trust visualizations uncritically, assuming they are accurate representations of facts, or distrust them due to a lack of understanding of how they are constructed.

\section{Methodology}

Prior work has looked at individual dimensions such as personality, cognitive factors, and visualization features to determine how each interacts with different mechanisms for evaluating trust~\cite{pandey2023you}. However, we know that humans are complex, and their processes for establishing, testing, and repairing trust when broken are likely to be nuanced. In this supplemental analysis, we use the data collected by Pandey et al. ~\cite{pandey2023you}, which defines five dimensions to measure trust: credibility, clarity, reliability, familiarity, and confidence. While the initial study ~\cite{pandey2023you} uses a multidimensional approach, including Kruskal-Wallis and post hoc Mann-Whitney tests, to explore the relationship between visual features and trust dimensions, this paper focuses more closely on the underlying relationships between these features through recursive partitioning. This approach enables us to more effectively isolate how these features move together or in tension with one another.

\subsection{Data Cleaning}

\begin{figure}[htp]
    \includegraphics[width= 0.48\textwidth]{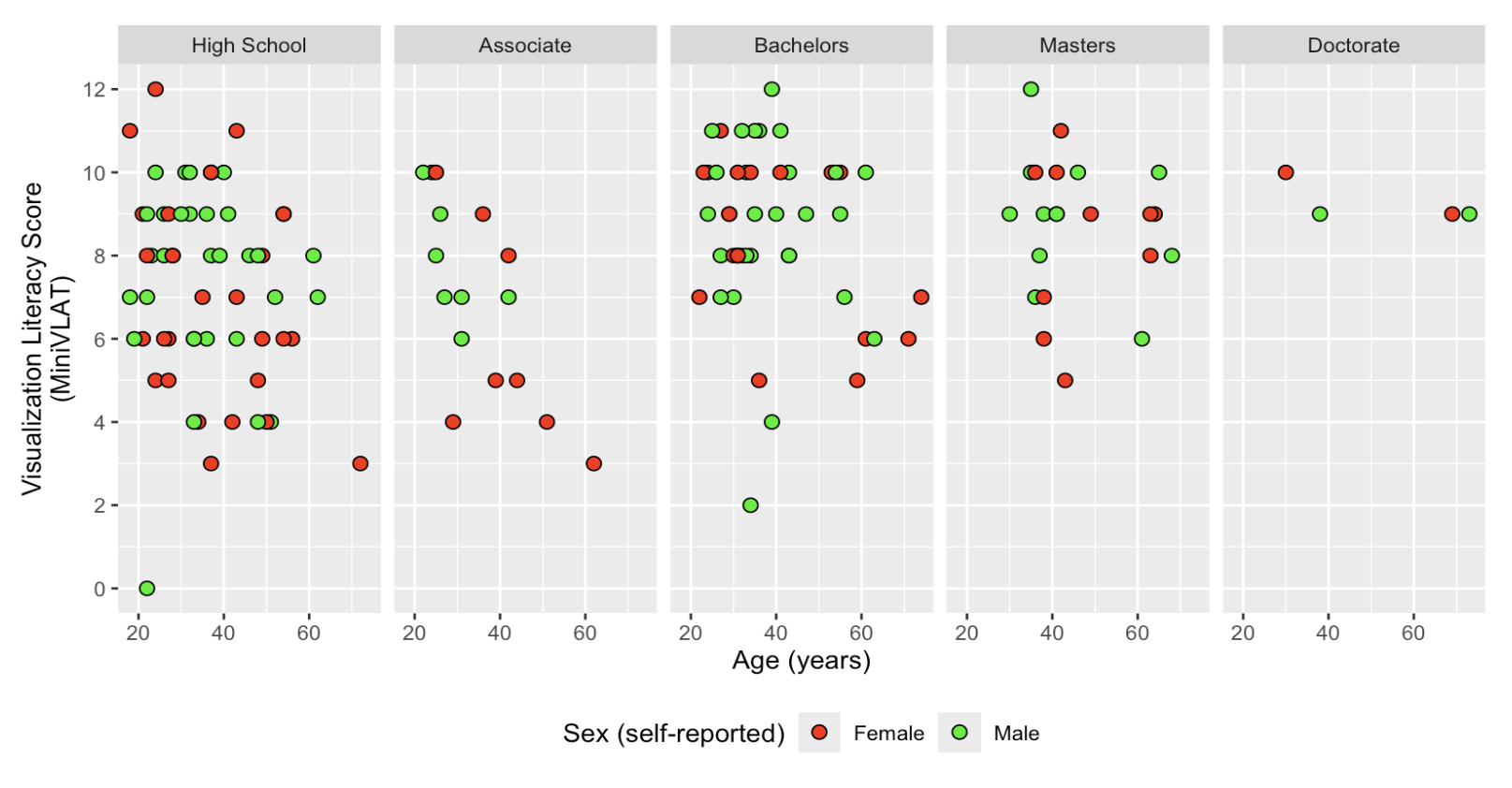}
    \caption{We observed a weak correlation between participants' visualization literacy and education. Samples in each education level were balanced for age and sex.}
    \label{fig:correlation}
\end{figure}

Prior to conducting our supplemental analysis, we carefully cleaned and prepared the collected data. We improved its descriptive quality by adding labels that specify the type of data visualization used and by converting the numeric scores from each Likert-scaled item to an ordinal response column ranging from 'strongly disagree' to 'strongly agree'. We also used correlation analysis to assess the independence of the features in question, so as to avoid undue bias toward features encoding redundant phenomena. We also ensured a balanced sample along the demographic dimensions of age and sex (see Fig.~\ref{fig:correlation}), after which we omit these dimensions from further analysis.


\subsection{Calculating Deviation-from-Average}

Because we are interested in the factors that influence trust, and the directions those influences steer the observer's assessment, we use \textbf{deviation from average trust} as our dependent variable rather than relying on the raw trust scores alone. We calculated the average scores in each dimension across all visualizations in the dataset and used that as a baseline by which to evaluate an individual visualization's deviation. Along each trust dimension, we then classified each image into three buckets: "higher agreement than usual (positive skew)," "lower agreement than usual (negative skew)," and "everything else." Higher and lower-than-usual agreement images were defined via the following formulas based on the overall agreement mean and standard deviation for each image:

\begin{itemize}
    \item \textbf{Higher agreement}: \text{agreement\_\(\mu\)} > \text{overall\_\(\mu\)} + \text{overall\_\(\sigma\)}
    \item \textbf{Lower Agreement}:  \text{agreement\_\(\mu\)} < \text{overall\_\(\mu\)} - \text{overall\_\(\sigma\)}
\end{itemize}

The cutoffs for these criteria resulted from experimentation. We then calculated the general response category for each image as follows:
\begin{itemize}
    \item \textbf{Higher general agreement}: \textbf{at least two dimensions} have higher trust and the rest are averages
    \item \textbf{Lower general agreement}: \textbf{at least two dimensions} have lower trust and the rest are averages
    \item \textbf{Mixed general agreement}: this image had both lower- and higher-than-average scores along different trust dimensions
\end{itemize}

Finally, we select the subset of images that across all categories had overall higher or lower trust to see what image attributes influenced that placement. The intuition is that when trust is skewed consistently in one direction (independent of the viewer), it could be attributed to an influence exerted by \textbf{endogenous} features of the image itself. Conversely, a mixed trust rating for the same image implies disagreement, which we hypothesize could be attributed to influence from \textbf{endogenous} factors related to the observer. To address the substantial skew towards strong agreement, we selected 125 images from each agreement level to ensure dataset balance. Columns unrelated to image attributes or individual characteristics were subsequently considered irrelevant and excluded to streamline the dataset.

We conducted a similar analysis along the individual axis, exploring which personal characteristics influence an individual's tendency to (mis)trust various visualizations. First, we grouped each trial by individual participants rather than images and calculated each individual's deviation from the sample average along each dimension. Interestingly, we observed that individuals tended to vary consistently across the five trust dimensions: 30\% of participants scored \textbf{either} higher or lower across any non-average dimensions, whereas just 12\% of participants skewed higher on one dimension and lower on another.

\subsection{Recursive Partitioning and Random Forests}

Recursive partitioning is a statistical modeling technique that repeatedly divides the data into smaller subsets based on specific variables, with the goal of making the resulting subsets as homogeneous as possible with respect to the dependent variable. Each split is chosen to best separate the data by minimizing variance within groupings (for regression problems) or maximizing purity within groupings (for classification problems). A random forest model constructs multiple recursive partition models during training and outputs the mode of the classes (for classification) or the mean/median prediction (for regression) across all models. This approach can provide more reliable models, in part because they are less susceptible to overfitting and undue influence from single, marginally-superior predictors than single decision trees.

\section{Results}
Our analysis identified both endogenous and exogenous factors that play a role in shaping people's trust in data visualizations. 
In each subsection, we will report the results of our ensemble methods (random forests); we will also include a representative tree whenever appropriate.

\begin{figure}[b!]
    \includegraphics[width=0.45\textwidth]{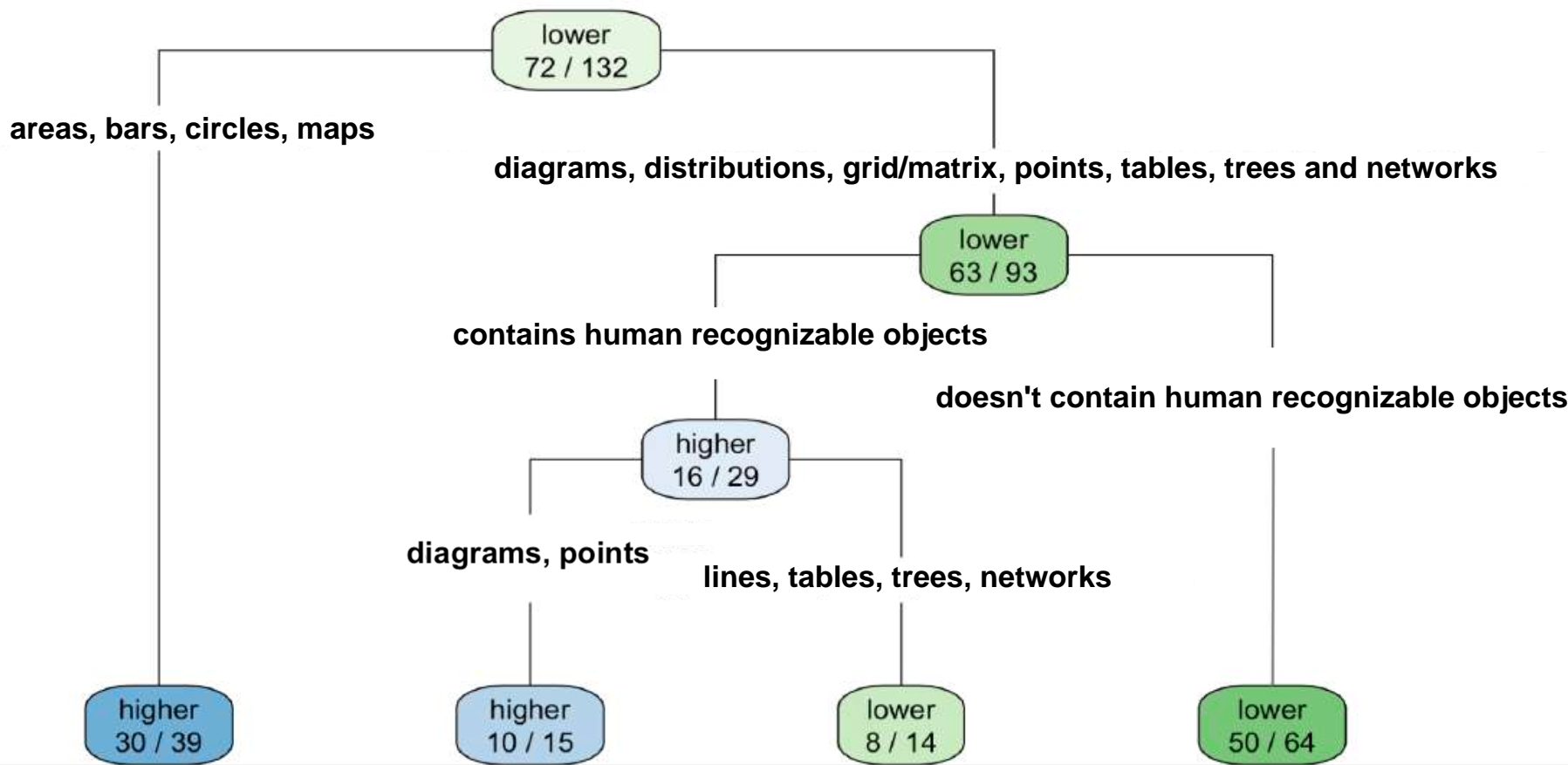}
    \caption{A representative tree depicting the influence of \textbf{endogenous} features on subjective trust ratings (higher or lower than average).}
    \label{fig:endogenous}
\end{figure}

\subsection{Endogenous Factors}
Both \textbf{visualization type} and the presence of \textbf{human-recognizable objects} rose to the front of the pack as significant predictors of trust. Specifically, visualizations employing areas, bars, circles, and maps generally engendered greater trust than those using diagrams, tables, trees, and grids. We also find that visualizations with human depiction tend to be more trusted by people. Furthermore, at the third level of the decision tree, we observed a potential trend where diagrams and points are associated with slightly higher levels of trust compared to lines, tables, trees, and networks. However since the difference is slight (10/15 for diagrams and points relative to 8/14 for lines and tables), we can not draw definitive conclusions from it.

\begin{figure}[t!]
    \includegraphics[width=0.48\textwidth]{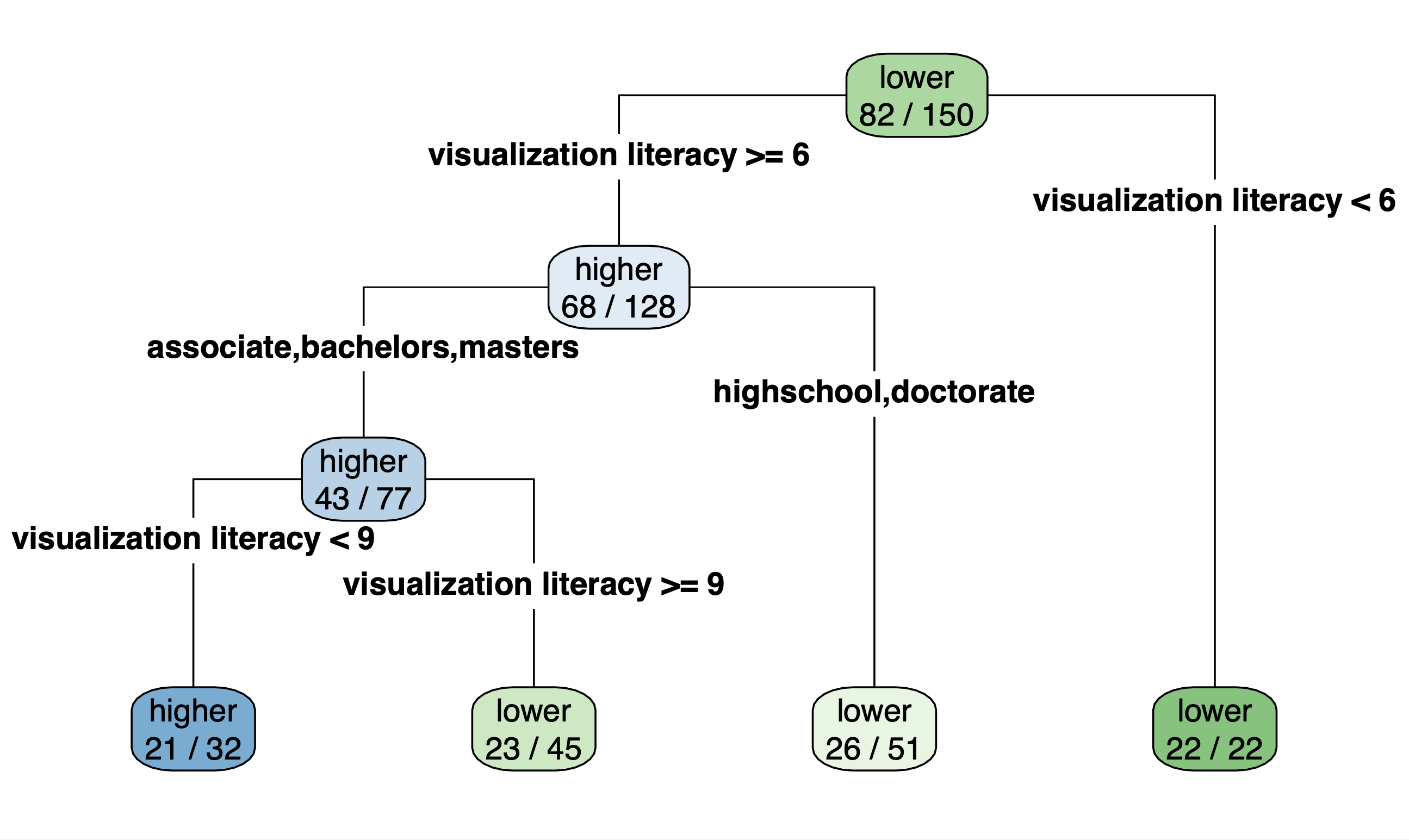}
    \caption{A representative tree depicting the influence of \textbf{exogenous} features on subjective trust ratings (higher or lower than average).}
    \label{fig:exogenous}
\end{figure}

\subsection{Exogenous Factors}
 Turning from features of the visualization to features of the observer, we find that \textbf{visualization literacy} (as measured by the MiniVLAT~\cite{pandey2023mini}) is the most significant predictor of deviation-from-average trust. Individuals with a score exceeding 6 typically exhibit higher trust compared to those scoring below 6, and this measure stands out as the primary predictor for both accuracy and purity. 
 
Figure~\ref{fig:exogenous} reveals an interesting bimodal distribution across \textbf{education}: those at the extremes of our sample (holding either high school diplomas or doctorate degrees) tended to demonstrate lower-than-average trust, whereas those with a moderate level of education (associate, bachelor's, or master's degree) tended toward higher-than-average trust \textit{especially} for those with a lower visualization literacy. This confounding highlights the dynamic nature of the forces at work: increased education may cause people to feel more confident in interpreting data visualizations, but when paired a persistently-low visualization literacy, the pendulum may swing into over-confidence.

\subsection{Endogenous and Exogenous Factors Combined}
Our analysis revealed consistent patterns in the importance of both \textbf{visualization literacy} and \textbf{visualization type} as top dimensions influencing the subjective trust rating of visualizations (as measured by their effects on both purity and accuracy). We also identified several factors whose influence appears to be tied to a specific dimension of trust.

\begin{itemize}

    \item Higher levels of \textbf{visualization literacy} and a higher \textbf{data-ink ratio} significantly enhance the \textbf{belief} that the visualized data is authentic. This suggests that when visualization-literate viewers encounter clean, no-frills visualizations, they are more likely to trust the authenticity of the data presented. \textbf{Bar, circle, grid, line, and map} types are particularly effective in bolstering this belief.
    \item In terms of \textbf{clarity}, visualizations that incorporate \textbf{human-recognizable objects} and use \textbf{circle or bar} chart formats are found to be clearer, more comprehensible, and more and memorable. These features aid in simplifying complex data, making it more accessible and thereby increasing viewer trust in the clarity of the information presented.
    \item Regarding \textbf{reliability}, specific formats like circle charts, tables, and maps are correlated with an increased belief that the data being depicted are reliable. These types of visualizations may be perceived as more structured and authoritative, enhancing their perceived reliability for conveying accurate information.
    \item \textbf{Familiarity} also plays a significant role, with visualizations in \textbf{bar, line, point, or table} formats being more recognizable to viewers, especially those with some higher education. This familiarity likely stems from frequent exposure to these formats in educational and professional settings, which translates into greater ease of understanding and a higher level of trust.
\end{itemize}

\section{DISCUSSION \& FUTURE WORK}

In light of our preceding findings, we can conclude that specific, widely applicable strategies can improve the clarity and trustworthiness of visualizations. The use of more straightforward data visualization formats, such as bar charts, and the inclusion of familiar objects can significantly enhance comprehension. Nevertheless, our research underscores the influence of variations within individual characteristics, such as data literacy and educational background, on the perceived trustworthiness of visual data representations. Therefore, addressing trust within data visualization necessitates efforts to boost data literacy levels through readily accessible data science education and resources. Moreover, it is essential to tailor visualizations to suit the audience’s diverse personalities and backgrounds.

It is also important to acknowledge the study’s limitations. Participants did not examine identical visualizations, and we did not verify the reliability of the visualizations that engendered trust. Thus, a key focus for our subsequent study involves conducting a pilot with controlled data. This will allow us to examine whether variations in individual attributes influence trust responses. While our current analysis provides insights into how these changes might impact trust, the proposed study will serve as a crucial validation, enabling us to assess the accuracy of our predictions. Considering the online nature of our data collection, participants inherently possess a degree of technological familiarity, which may be influenced by factors such as age, education, and data literacy. In future studies, we aim to design more inclusive experiments by considering a wider age range and accounting for varying levels of technological proficiency. Results from this investigation will inform design recommendations for data visualizations, contributing to refining practices in this domain.

\acknowledgments{
This work is partly supported by the National Science Foundation under Grant No. IIS-2142977.
It is also based upon work done, in whole or in part, in coordination with the Department of Defense (DoD). Any opinions, findings, conclusions, or recommendations expressed in this material are those of the author(s) and do not necessarily reflect the views of the DoD and/or any agency or entity of the United States Government.}

\bibliographystyle{abbrv-doi-hyperref}

\bibliography{template}

\appendix 

\end{document}